\documentclass[aip,jap,reprint]{revtex4-1}
\usepackage{graphicx}

\begin{document}


\title{Role of material properties and mesostructure on dynamic deformation and shear instability in Al-W granular composites} 

\author{K.L. Olney}
\affiliation{Department of Mechanical and Aerospace Engineering, University of California at San Diego, La Jolla, California 92093-0411, USA}
\author{P.H. Chiu}
\affiliation{Materials Science and Engineering Graduate Program, University of California at San Diego, La Jolla, California 92093-0411, USA}
\author{C.W. Lee}
\affiliation{Department of Mechanical and Aerospace Engineering, University of California at San Diego, La Jolla, California 92093-0411, USA}
\author{V.F. Nesterenko}
\affiliation{Department of Mechanical and Aerospace Engineering, University of California at San Diego, La Jolla, California 92093-0411, USA}
\affiliation{Materials Science and Engineering Graduate Program, University of California at San Diego, La Jolla, California 92093-0411, USA}
\author{D.J. Benson }
\affiliation{Department of Structural Engineering, University of California at San Diego, La Jolla, California 92093-0085, USA}


\date{\today}

\begin{abstract}
Dynamic experiments with Al-W granular/porous composites revealed qualitatively different behavior with respect to shear localization depending on bonding between Al particles. Two-dimensional numerical modeling was used to explore the mesomechanics of the large strain dynamic deformation in Al-W granular/porous composites and explain the experimentally observed differences in shear localization between composites with various mesostructures. Specifically, the bonding between the Al particles, the porosity, the roles of the relative particle sizes of Al and W, the arrangements of the W particles, and the material properties of Al were investigated using numerical calculations. It was demonstrated in simulations that the bonding between the “soft” Al particles facilitated shear localization as seen in the experiments. Numerical calculations and experiments revealed that the mechanism of the shear localization in granular composites is mainly due to the local high strain flow of “soft” Al around the “rigid” W particles causing localized damage accumulation and subsequent growth of the meso/macro shear bands/cracks. The “rigid” W particles were the major geometrical factor determining the initiation and propagation of “kinked” shear bands in the matrix of “soft” Al particles, leaving some areas free of extensive plastic deformation as observed in experiments and numerical calculations.
\end{abstract}


\maketitle 
\section{Introduction}
The dynamic behavior of granular/porous reactive materials has attracted significant attention due to its possible practical applications: reactive structural components, reactive fragments, etc.\cite{Ames,davis,holt} The performance requirements introduce challenging fundamental problems such as combining high strength with ability to undergo controlled bulk disintegration producing small sized reactive fragments.  

The material structure which can meet this and other requirements should be tailored and optimized at the mesoscale to produce the desirable mechanical properties while still facilitating the release of chemical energy. Mesostructural parameters like particle size and morphology can affect the strength and shock-sensitivity. This can be seen in pressed explosives\cite{Siv,Balzer} and in reactive materials like Al-PTFE composites.\cite{Mock} Additionally, mesoscale features like force-chains in granular energetic materials may also serve as ignition sites.\cite{foster,Bard,Roes}

Quasi-static, Hopkinson bar, and drop-weight experiments were performed for PTFE-Al-W and for Al-W composites\cite{Cai1,Cai2,Herbold,Cai3,Cai4,Cai5,Herbol1,Dymat,phaip} with different particle sizes of W, porosity, and morphology. W particles were used to increase sample density and generate the desirable mode of disintegration of the sample into small sized debris. Multi-material Eulerian hydrocode simulations of the dynamic tests for the various types of samples were used to elucidate the observed experimental results.\cite{Cai2,Herbol1,phaip}

The complexity of the dynamic behavior of granular/porous materials is caused by the significant role of the mesoscale parameters including the particle sizes of components, the formation of force chains, the morphology of particles, and the bonding between particles. In these materials, shear localization and fracture can be delayed by strain hardening mechanisms such as compaction which results in porosity reduction.\cite{Dymat,phaip} At the same time, mesoscale fracture of brittle particles can act as a “softening” mechanism increasing macroshear instability.\cite{NesMey,Shih} In this paper, the dynamic behavior of Al-W granular/porous composites, their susceptibility to shear localization, and their subsequent fracture at large strains was investigated using experiments and numerical simulations. Specifically, the bonding between the Al particles, the initial porosity, the relative particle sizes of Al and W, the arrangements of the W particles, and the constitutive behavior of Al was examined in Al-W granular/porous composites.
\section{Experiments}
\begin{figure}
\includegraphics{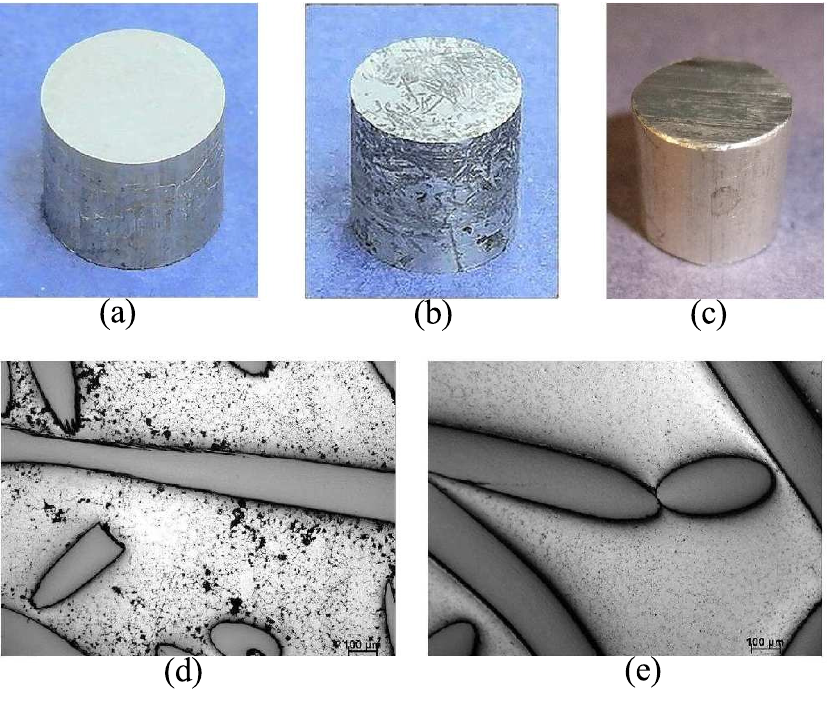}
\caption{\label{fig:1} (Color online) Initial CIPed (a) and CIPed+HIPed Al-W samples (b) and a sample of  solid 6061-T6 (c). Sample diameters and heights are given in Table \ref{tab:4}. Mesostructures of  CIPed and CIPed+HIPed samples before dynamic deformation are presented in (d) and (e) respectively.  }
\end{figure}

\begin{figure}
\includegraphics{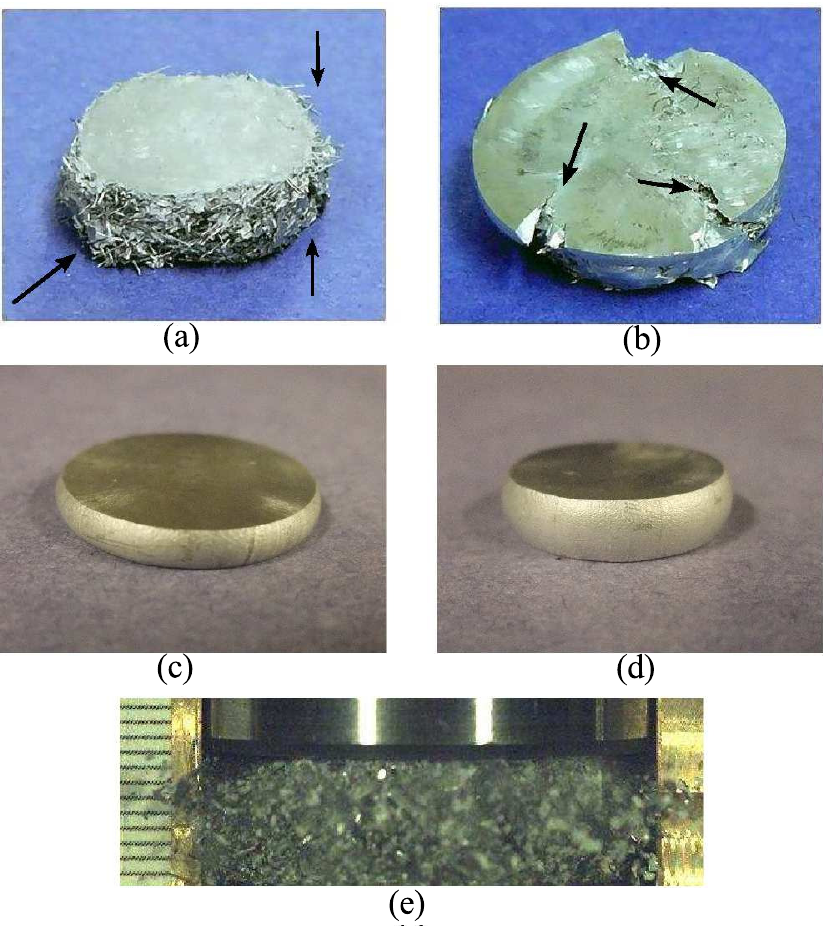}
\caption{\label{fig:2} (Color online) Deformed shapes of the CIPed (a) and CIPed+HIPed samples (b). For comparison, the deformed annealed Al-6061-T6 samples and Al-6061-T6 are presented correspondingly in (c) and (d). A high speed camera photo (e) captures the behavior of Al fragments as they are ejected from the CIPed sample during a dynamic test. Arrows are added to show areas where the Al particles on the outer area of the CIPed sample have been ejected (a) and the shear bands that form in the CIPed + HIPed samples (b).    }
\end{figure}

\begin{figure}
\includegraphics{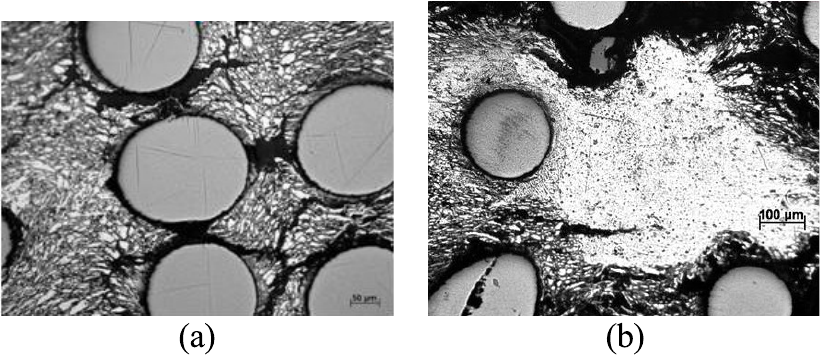}
\caption{\label{fig:3} (Color online) Microstructure of deformed CIPed+HIPed samples (a) and (b) illustrating a region of largely undeformed Al particles neighboring heavily deformed and fractured areas of Al particles with different arrangements of less deformed W rods.   }
\end{figure}

High density Al-W granular/porous composites were prepared from an elemental powder of Al (Alfa Aesar,-325 mesh), and W wires (A-M System, diameter of 200 $\mu$m and length of 4mm) using (a) cold isostatic pressing (CIPing) and (b) CIPing followed by vacuum encapsulation and subsequent hot isostatic pressing (HIPing) to create metal bonding between Al particles. All samples had the same mass ratios of the Al and W components (23.8\% Al and 76.2\% W, by weight, corresponding to a volume ratio of 69.0\% Al and 31.0\% W) with a theoretical density of 7.8 g/cm3. Mixtures of Al powders and W (short wires) were placed in a cylindrical stainless steel mandrel with moving pistons and encapsulated in a rubber jacket providing axial loading during pressurization in the CIP chamber. This allowed the preparation of samples with the exact sizes and cylindrical shape that were necessary for the strength measurements. All samples were CIPed at 345 MPa under room temperature for 5 min. Subsequent HIPing for some samples were carried out at 200 MPa at a temperature of 500 degree Celsius with a soaking time 20 min. The density of the samples was measured by the hydrostatic method. The samples had an average density of 6.8 g/cm$^3$ after CIPing and 7.5 g/cm$^3$ after CIPing and HIPing. Their dynamic behavior and fracture was investigated at a strain rate of 1000 1/s under drop weight tests (nominal velocity of the drop weight was 10 m/s) using a high velocity DYNATUP model 9250HV with an in-house modified anvil supporting the sample. The undeformed CIPed only, CIPed+HIPed, and solid Al6061-T6 samples are presented in Fig. \ref{fig:1}. After the dynamic tests, shear macrocracks were well developed in the CIPed+HIPed samples (see Fig. \ref{fig:2} (b)), while no shear macrocracks were observed in the CIPed only samples (see Fig. \ref{fig:2} (a)) at similar global strains. In the CIPed only sample, Al particles were ejected from the outer regions of samples leaving areas of interconnected and mechanically locked W wires (see Fig. \ref{fig:2} (a) and (e)). The microscopic images of the deformed samples representing areas at the vicinity of macroscopic shear localization, plastic flow and deformation of W and Al particles are presented in Fig. \ref{fig:3}. It is evident that the bonding between the “soft” Al particles facilitated shear localization in CIPed+HIPed samples. As a result of localized plastic flow in the matrix of Al particles we have observed regions of heavily deformed and practically undeformed Al particles. 
 
For comparison, dynamic tests were performed with “as is” Al6061-T6 and annealed Al6061-T6 (temperature 425C, 2.5 hours in vacuum) cylindrical samples under similar conditions of impact and similar or larger final strains in comparison with CIPed or CIPed+HIPed samples (see Fig. \ref{fig:2}). The “as is” Al6061-T6 and annealed Al6061-T6 samples did not exhibit signs of shear localization at this level of strains. 

\begin{table}
\caption{\label{tab:4} Initial dimensions of experimental sample}
\begin{ruledtabular}
\begin{tabular}{lrr}
Sample&Height [cm]&Diameter [cm]\\
\hline
Ciped only&1.45&1.58\\
Ciped + Hiped&1.39&1.54\\
Solid Al 6061-T6&1.25&1.18\\
\hline
\end{tabular}
\end{ruledtabular}
\end{table}
\begin{table}
\caption{\label{tab:1} Particle sizes and distributions}
\begin{ruledtabular}
\begin{tabular}{lrr}
&Particle Diameter&\% of Material\\
&$\mu$m&\\
\hline
-325 mesh&32& 50\\
&15&40\\
&4&10\\
\hline
Fine&40&-\\
\hline
Coarse&200&-\\ 
\end{tabular}
\end{ruledtabular}
\end{table}
\begin{table}
\caption{\label{tab:2} Material volume fractions and porosity in each of the simulated samples}
\begin{ruledtabular}
\begin{tabular}{lrrr}
Sample in &Volume&Volume&\% Initial\\
Coresponding&Fraction&Fraction&Porosity\\
Figure&Al&W&\\
\hline
1.a&0.72&0.28&8.3\\
1.b&0.74&0.26&0.0\\
2&0.67&0.33&8.5\\
\end{tabular}
\end{ruledtabular}
\end{table}
\section{Numerical Modeling and Discussion}
A number of sample properties were examined to see how they influenced shear instability and localization during a dynamic test. The role of the metallic bonding was examined in an attempt to understand a difference between dynamic behavior of CIPed samples (no metallic bonding between Al particles) and CIPed+HIPed samples (metallic bonding between Al particles). The roles of the initial porosity, the relative particle size of Al and W, the initial arrangement of the W particles, the confinement, and the constitutive behavior of the Al were also explored to see their effect on the shear instability in the sample. A two-dimensional Eulerian hydocode\cite{Benson} was used to simulate the behavior of the samples during the drop weight tests. Due to the smallness of the particles relative to the global sample size, small representative elements of the microstructure were used in all the simulations.  
\subsection{The initial particle arrangements, the constitutive model, and the boundary conditions}
\begin{figure}
\includegraphics{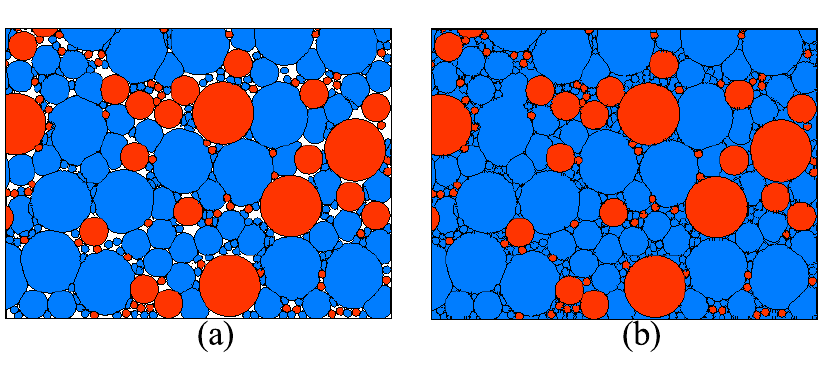}
\caption{\label{fig:4}  (Color online) Initial arrangements of the Al (blue) and the W (red) particles in the Al-W granular/porous representative samples with similar sized Al and W particles corresponding to -325 mesh (see Tables \ref{tab:1} and \ref{tab:2}) with .3\% (a) and 0\% (b) initial porosity. Representative element sample sizes are 0.02 cm (horizontal) by 0.015 cm (vertical). These samples were used for both the bonded and the unbonded simulations.  }
\end{figure}

\begin{figure}
\includegraphics{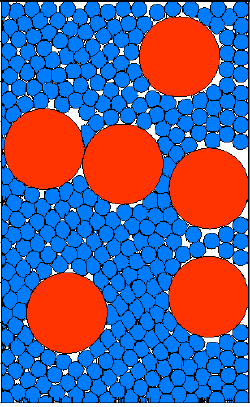}
\caption{\label{fig:5} (Color online) Initial arrangement of the Al (blue) and the W (red) particles in the Al-W granular/porous sample with fine Al (40 μm) and coarse W (200 μm) particles, and a porosity of 8.5\%. This arrangement was used both for the bonded and the unbonded Al particles. Sample size is 0.0625 cm (horizontal) by 0.1 cm (vertical).	   }
\end{figure}

\begin{figure}
\includegraphics{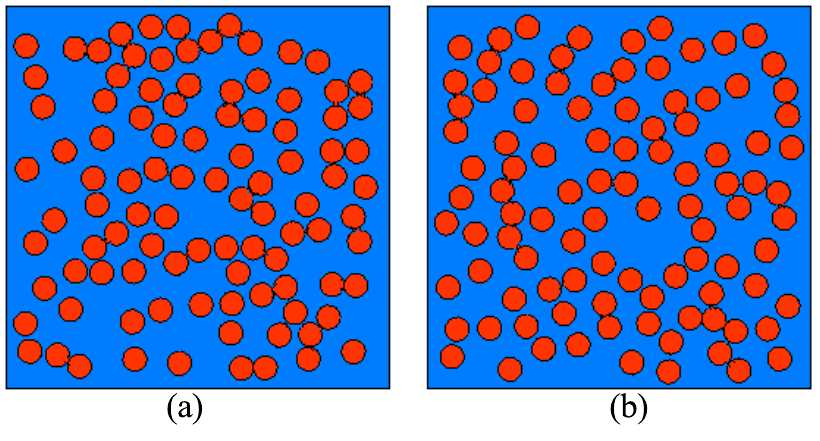}
\caption{\label{fig:6} (Color online) The initial arrangement of the 200 μm diameter W (red) particles embedded in a fully dense (0\% porosity) Al (blue) matrix. Samples have sizes of 0.3 cm (horizontal) by 0.3 cm (vertical). Both samples have the same volume fractions of Al and W of 0.70 and 0.30 respectively.    }
\end{figure}

The initial particle arrangements of W and Al used in the numerical simulations are shown in Fig. \ref{fig:4} through Fig.\ref{fig:6}. The description of the particles sizes for both the Al and W are presented in Tables \ref{tab:1} and \ref{tab:2} . Samples in Fig. \ref{fig:4} (a) and (b) were used to investigate the role of porosity and metallic bonding on the shear instability and formation of the shear bands in both the bonded (CIP+HIP) and unbonded (CIP only) samples with similar sized Al and W particles. The sample in Fig. \ref{fig:5} was used to investigate how relative sizes of W and Al affected the shear instability in both the bonded and unbonded samples. Additionally, the sample in Fig.\ref{fig:5} was used to examine cases with “confinement” boundary conditions as well as the variations for simulations looking at the role of constitutive behavior in Al. Samples in Fig. \ref{fig:6} were used to investigate the role of initial mesostructure on the shear instability in the bonded case. 

A standard Johnson-Cook with failure\cite{Jcook} material model and the Mie-Grüneisen equation of state used for both the Al and the W in the simulations. Failure in the material is determined through the equivalent plastic strain as shown in Ref. \onlinecite{Jcook}. The material constants used in this paper are presented in Table \ref{tab:3}. 

\begin{table}
\caption{\label{tab:3} Consitutive model and equation of state material parameters}
\begin{ruledtabular}
\begin{tabular}{lrr}
&Al 6061-T6\cite{Holm,Stein}&W\cite{Holm,Wester} \\
\hline
Johnson-cook&&\\
\hline
$\rho$ [gm cm$^{-1}$]&2.7&16.98 \\
G [Mbar]&0.26 &1.24 \\
A [Mbar]&2.24$\cdot$10$^{-3}$&1.506$\cdot$10$^{-2}$\\
B [Mbar]&1.114$\cdot$10$^{-3}$&1.765$\cdot$10$^{-3}$\\
n&0.42&0.12\\
c&2.0$\cdot$10$^{-3}$&1.60$\cdot$10$^{-3}$\\
m&1.34&1.0\\
C$_p$ [J gm$^{-1}$ K$^{-1}$]&0.89&0.13\\
T$_{melt}$ [K]&930&1728\\
T$_{room}$ [K]&300&300\\
D$_1$&-.077&0.0\\
D$_2$&1.45&0.33\\
D$_3$&-0.47&-1.5\\
D$_4$&0.0&0.0\\
D$_5$&1.6&0.0\\
\hline
Mie-Grüneisen&&\\
\hline
c$_0$ [cm $\mu s^{-1}$]&0.52&0.40 \\
s$_1$&1.4&1.24\\
s$_2$&0.0&0.0\\
s$_3$&0.0&0.0\\
$\gamma_0$&1.97&1.67\\
\hline
\end{tabular}
\end{ruledtabular}
\end{table}

\begin{figure}
\includegraphics{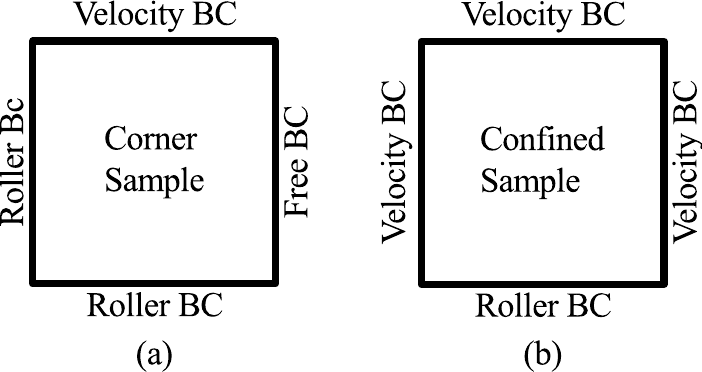}
\caption{\label{fig:7}  Boundary conditions imposed on (a) corner samples and (b) internal confined samples.   }
\end{figure}

To simulate the drop weight test, a kinematic boundary condition with constant downward velocity was imposed on the top boundary corresponding to the impact speed of the falling weight in the experiment (10 m/s). For the cases where the small representative element is near the outer edge of the large cylinder, the side wall boundary conditions shown in Fig. \ref{fig:7} (a) were used. In cases where the small representative element is near the middle of the large cylinder, the side wall boundary conditions shown in Fig. \ref{fig:7} (b) were used to account for the confinement. A variety of confinement conditions were considered in this paper and will be detailed during the discussion of Fig. \ref{fig:13}. With the confinement boundary conditions, the main concern is keeping the correct description of the global deformation of the sample, but due to the artificial conditions, the agreement with global stresses is sacrificed. This “confined” geometry produces compressive stresses in particles leading to elevated pressures in the sample.

\subsection{The role of bonding and initial porosity in samples with similar sized Al and W particles (-325 mesh)}
\begin{figure}
\includegraphics{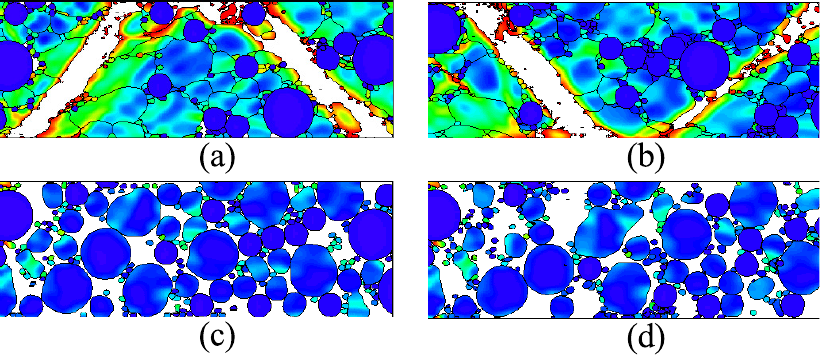}
\caption{\label{fig:8} (Color online) Role of initial porosity and bonding between particles. Deformed mesostructures of -325 mesh Al and W particles with the initial arrangement of Al and W particles shown in Fig. \ref{fig:4} (a) and (b). Damage is plotted to highlight the shear instability and shear bands in the sample. All samples shown at global strain of 0.45 with in itial particle arrangement presented in:\\
(a) Fig. \ref{fig:4} (a) with bonded particles (CIPed + HIPed).\\
(b) Fig. \ref{fig:4} (b) with bonded particles (CIPed + HIPed).\\
(c) Fig. \ref{fig:4} (a) without bonded particles (CIPed only).\\
(d) Fig. \ref{fig:4} (b) without bonded particles (CIPed only).}
\end{figure}

\begin{figure}
\includegraphics{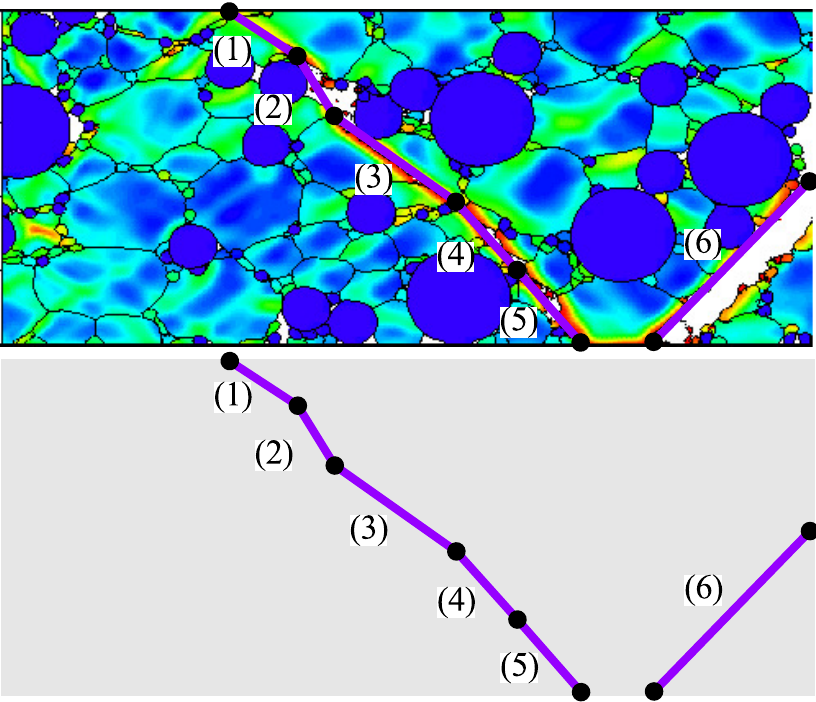}
\caption{\label{fig:9}  (Color online) Effect of rigid W particles on the “kinking” of the shear bands in the Al matrix with the W and the Al -325 mesh sized particles with the initial mesostructure illustrated in Fig. \ref{fig:4} (a).  Black dots show the kink locations of the shear bands.  The angles of the piecewise shear bands are 32 (1), 58 (2), 34 (3), 47 (4), 48 (5), and 45 (6) degrees.    }
\end{figure}

The results of the numerical simulations corresponding to the initial mesostructures shown in Fig. \ref{fig:4} are presented in Fig. \ref{fig:8} corresponding to the deformed sample with a global strain of 0.45. The sample with the bonded particles and initial porosity of 8.3\% began developing shear instabilities at the mesoscale once the in situ densification that occurred at the early stages of deformation removed a majority of the porosity from the sample (the initial average porosity was 8.3\% and the porosity when shear instability started was 3.2\%). The bonded sample with zero initial porosity began developing shear instabilities on the mesoscale within the first 5\% global strain. The global shear band began developing around 0.25 global strain in the case with 8.3\% initial porosity and around 0.20 global strain in the sample with zero initial porosity. This behavior is similar to observations in the PTFE-Al-W mixtures and in the experiments with the Al-W powder mixture.\cite{Cai2,Herbold,Herbol1,Dymat,phaip} The global shear localized zones and the subsequent cracks developed at an approximate 45 degree angle but was “kinked” by the rigid W particles resulting in the initial shear band forming in a range of angles from 36 to 50 degrees (see Fig. \ref{fig:9}). The samples in Fig. \ref{fig:8} (a) and (b) have a relatively large percentage of small W particles, providing some geometrical “homogenization” of the mixture. As a result, the shear bands do not significantly deviate from this 45 degree angle. In both samples, the mesoscale shear instabilities continue to grow until one begins to dominate (at approximately 0.25 global strain) and a global shear band develops, which in turn leads to the growth of a macro crack. While this simulation only looks at a small representative element along the edge, the experimental CIP+HIP samples exhibit the same behavior (see in Fig. \ref{fig:2} (b)).

The corner samples that have unbonded particles, simulating the CIPed only material, are presented in Fig. \ref{fig:8} (c) and (d). These samples do not develop shear instabilities like the bonded samples with identical initial mesostructures. In the unbonded samples, the Al and W particles rearrange themselves during the dynamic deformation, effectively blocking shear instabilities from forming. Additionally, due to the free boundary on the right hand side, the sample undergoes bulk disintegration. Due to the two-dimensionality of the simulation and the lack of friction in the simulation, the W particles do not offer much resistance to the movement of the Al out of the right boundary. A similar behavior can be seen in the experimental samples where the Al on the edge is ejected from the sample, disintegrating into agglomerates of Al consisting of 5-30 initial sized Al particles. This ejection of the Al particles from the network of the W wires adjacent to the free boundary was observed during the dynamic loading in experiments using a high speed camera (see Fig. \ref{fig:2} (e)).

Initial variations in porosity (0\% or 8.35\%) in the bonded sample altered the orientation of the mesoscale shear bands despite the practically identical initial mesostructure of the W particles. In the sample with porosity, the pores are bulk distributed with sizes of approximately 4 microns in diameter. This bulk-distributed porosity allows for small movements of the W particles at the initial stages of deformation due to pore closure. The shear instabilities were observed nucleating during this early stage of the deformation. The small alterations in the W particle arrangement at this early stage therefore caused a large deviation in shear bands that developed during later stages of the deformation. The unbonded samples with and without porosity both showed the same characteristic of particle rearrangement blocking shear band formation and bulk disintegration.

\subsection{The role of bonding and initial porosity on samples with fine (40 micron) Al and coarse (200 micron) W particles}
\begin{figure}
\includegraphics{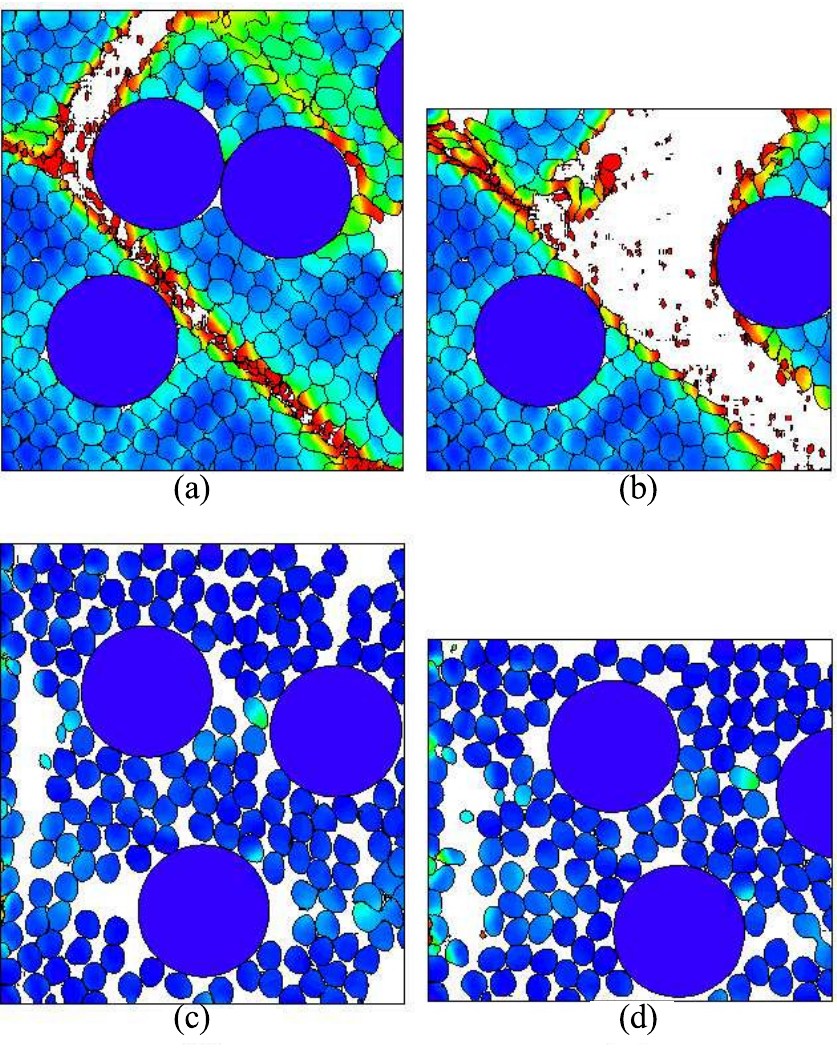}
\caption{\label{fig:10} (Color online) Role of relative particle sizes of W and Al in bonded (a), (b) and unbounded samples (c), (d) at two different global strains of 0.30 (a), (b) and 0.45 (c), (d). The initial mesostucture is shown in Fig. \ref{fig:5}. Damage is plotted to highlight the shear instability and the shear band development.   }
\end{figure}

Numerical calculations exploring how differences in the relative sizes between the W and Al particles (see the initial mesostructure in Fig. \ref{fig:5}) effect the development of shear instabilities and the formation of shear bands/cracks are shown in Fig.\ref{fig:10}. The W particles in the numerical simulations have a diameter of 200 microns, the same diameter as the W rods used in the experimental samples shown in Fig. \ref{fig:1} (a) and (b). The Al particles have a diameter of 40 microns, the size of the larger particles in the -325 mesh particle size used in the experimental sample. The boundary conditions are the same as those described in Fig. \ref{fig:7} (a) with a constant downward velocity of 10 m/s on the top boundary.

In the bonded sample simulating the CIPed+ HIPed material (Fig. \ref{fig:10} (a) and (b)), shear instabilities begin developing after the pores close like the porous sample with similar sized Al and W particles shown in Fig. \ref{fig:8} (a). Mesoscale shear bands develop at approximately 0.25 global strain. The shear band that forms is heavily influenced (guided) by the W particles in its path. When comparing the shear bands in Fig. \ref{fig:10} with those in Fig. \ref{fig:8}, the W particles that are larger relative to the Al have a much greater influence on directing and altering the path of the shear bands. The larger W particles require the shear instability to circumvent a larger radius of the relatively rigid W particle to connect with instabilities that form in the Al on the other side of the W particle. In Fig. \ref{fig:10} (a) the shear band can be clearly seen circumnavigating the W particle in the upper central region, causing the shear band to turn almost 90 degrees from one side of the particle to the other. The shear band in the lower section of the same figure is directed between two adjacent W particles. Due to the larger relative size of the W to the Al particles, the heterogeneous nature of the sample is increased in comparison to the previous samples in Fig. \ref{fig:8}, where the small W particles homogenized the mesostructure. This creates areas of mostly undeformed Al to appear in regions away from W particles and areas of deformed Al to appear in regions near W particles. Based on the simulations, the path of the shear bands closely follows the W particles in CIPed+Hiped material. This behavior is in agreement with the experiments (see Fig. \ref{fig:3}).

In the unbonded sample simulating the CIPed only material in Fig. \ref{fig:10} (c) and (d), the bulk distributed rearrangement of the Al and W particles effectively block the development of shear bands. This rearrangement is similar to that of the particles in Fig. \ref{fig:8} (c) and (d). Like the previous case due to limitations in the two-dimensionality of the simulations some of the three-dimensional W fiber effects cannot be reproduced. However, the trends seen in this two-dimensional simulation can also be seen at the edge of the experimental sample where the Al is ejected from the edge of the cylinder during the dynamic test as shown in Fig. \ref{fig:2} (a) and (e).

\subsection{The relationship between sample strength and porosity in corner samples}
\begin{figure}
\includegraphics{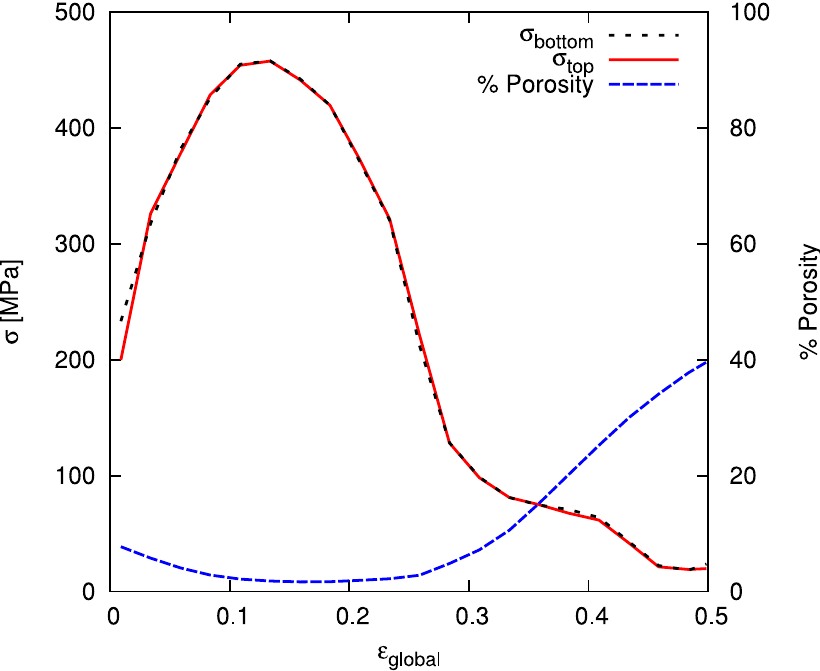}
\caption{\label{fig:11}  (Color online) Average engineering stress versus the global strain at the top boundary (solid red line) and at the bottom boundary (dashed black line) of the corner sample corresponding to initial mesostructure presented in Fig. \ref{fig:5} and velocity of impact 10 m/s. The porosity in the sample versus global strain is also plotted (blue dashed and dotted line).  }
\end{figure}

The engineering stress at the upper and lower boundaries versus the global strain for the simulation of bonded sample described in Fig. \ref{fig:10} is shown in Fig. \ref{fig:11}. The solid red curve and the dashed black curve show the average engineering stress on the top and bottom boundaries respectively and are nearly identical except for the first 4\% strain. Based on this, the sample can be seen as undergoing a quasi-static deformation. In addition to the stress, the percentage of porosity in the sample was plotted to show the relationship between the porosity and sample strength. It can be seen that stresses in the sample increased until a global strain of 0.12 was reached. The maximum of stress corresponds to the minimum porosity due to in situ densification during the initial stage of deformation. After the densification stage (corresponding to a range of global strain 0.12-0.29), the stress begins to drop as the local shear bands begin forming and the global shear band develops. Later (0.30 global strain) macrocracking occurred resulting in a decrease in strength accompanied by a rapid increase of porosity. This behavior is similar to the behavior observed in.\cite{Cai2,Herbold,Herbol1,Dymat,phaip} The two-dimensional simulation of the unbonded material, due to the material rearrangement, had near zero engineering stress on the top and bottom boundary and was not included. The comparable strength of bonded and unbounded samples in experiments is mostly due to interconnected W wires that cannot be accounted in these two-dimensional simulations.

\subsection{The role of initial arrangement of W particles in CIP+HIP samples (bonded) with zero initial porosity}
Current processing techniques create samples where the W fibers are initially placed in a random fashion. Multiple “randomized” samples were created for numerical analysis to understand how different experimental realizations of this randomized W fiber placement effects the formation of shear instabilities during dynamic deformation. In two characteristic samples with initial mesostructure presented in Fig. \ref{fig:6}, the W particles were randomly placed in an Al matrix such that the volume fractions of W and Al in each sample were identical. The samples were then subjected to boundary conditions described in Fig. \ref{fig:7} (a) with a constant 10 m/s downward velocity. All samples had zero initial porosity to remove mesostructural changes due to void closure as seen in simulations described in Fig. \ref{fig:8}. Results of the two characteristic simulations are shown in Fig. 12, demonstrating variations in the shear instabilities and shear band development while highlighting the similarities that the samples share.

\begin{figure}
\includegraphics{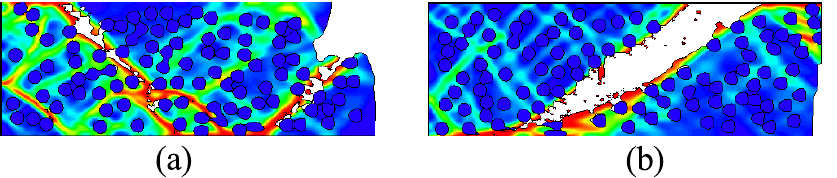}
\caption{\label{fig:12}(Color online) Role of initial mesostructure of the W particles on the shear localization and the subsequent fragmentation with zero initial porosity. The deformed samples correspond to the initial W particles arrangements shown in Fig. \ref{fig:6} :\\
(a) at 0.5 strain corresponding to Fig. \ref{fig:6} (a) \\
(b) at 0.5 strain corresponding to Fig. \ref{fig:6} (b) 
     }
\end{figure}

Both samples in Fig. \ref{fig:12} have damage plotted to highlight the shear band formation, with red corresponding to fully damaged Al. It is obvious from the comparison of the deformed samples in Fig. \ref{fig:12} (a) and (b) that changes in the initial mesostructure (see Fig. 6 (a) and (b)) greatly alter the location of the shear band despite the same volume content of Al and W and the same size of W particles. 

While both samples in Fig. \ref{fig:12} have very different shear band locations and modes of fracture, there are a number of similar characteristics that the samples exhibit. 

First, both samples develop shear instabilities and shear bands at approximately the same global strain of 0.2. It should be mentioned that in these samples with zero initial porosity, the global strains corresponding to well developed shear localization are similar to the corresponding global strains in the initially porous samples (see Fig. \ref{fig:8} and Fig. \ref{fig:10}). 

Second, numerous local shear instabilities form due to the localized, high strain flow of softer Al around the harder W particles. These local instabilities link with other local shear instabilities in close proximity creating localized shear bands. These bands join with other localized shear bands until one global shear band transverses the entire sample and becomes the dominant macroshear band. This macroshear band has a propensity to form at a 45 degree angle, with its path locally altered by W particles, and spans the entire sample before subsequent global shear bands are able to form. This demonstrates that the relatively rigid W particles initiate the shear instability in these granular composites, enhancing the localized high strain plastic flow of Al around the W particles. Local plastic strain in the Al around the W particles was 3 to 4 times higher than the plastic strain in the surrounding Al thus facilitating localized damage and subsequent global shear localization within the Al matrix. Separate experiments and numerical simulations with cylindrical samples made of as is Al6061-T6 and annealed Al6061-T6 did not reveal shear localization at similar or larger global strains (see Fig. \ref{fig:2} (c) and (d)). 

Third, the W particles are largely responsible in dictating the path of instabilities. This can be seen clearly in the bottom center region of Fig. \ref{fig:12} (a), where the shear band develops two branches that circumvent a clump of W particles and reconvene on the bottom boundary. Additionally, in the top central region of Fig. \ref{fig:12} (a), a closely packed clump of W particles block shear instabilities from developing in this region. This may be due to a similar effect as seen in Fig. \ref{fig:8} (a) and (b) in addition to Fig. \ref{fig:10} (a), where having the W particles initially aggregated into a close pack causes the shear instability to move around the relatively rigid mass into the softer area of the surrounding composite. The same effect can be seen in all areas of the samples in Fig. \ref{fig:12} where the W particles group in very close proximity to each other. 

Finally, due to the random nature of the sample, some of the W particles form angular chains with channels of Al between them facilitating the growth of shear instabilities and shear bands along the channel. The sample in Fig. \ref{fig:12} (b) exemplifies this with W particles creating a channel structure spanning the entire sample going diagonally from the lower left corner to the upper right corner. This channel created a favorable path for the local instabilities to follow and eventually lead to the formation of the shear band in the sample. W particles “flow” along the shear band to help form these angular chains. Rearrangement of the rigid particles inside high strain shear flow was observed in explosive compaction on interfaces between particles with high strain shear deformation.\cite{Nbook} In Fig. \ref{fig:12} (a) a distinct global chain structure of W particles is not present like in Fig. \ref{fig:12} (b), but rather a series of shorter chain structures in the left and the upper right region of the sample. This lack of long chains may be the reason that the sample accumulated more bulk-distributed damage in comparison to Fig. \ref{fig:12} (b). 

\subsection{The role of kinematic confinement conditions }
\begin{figure}
\includegraphics{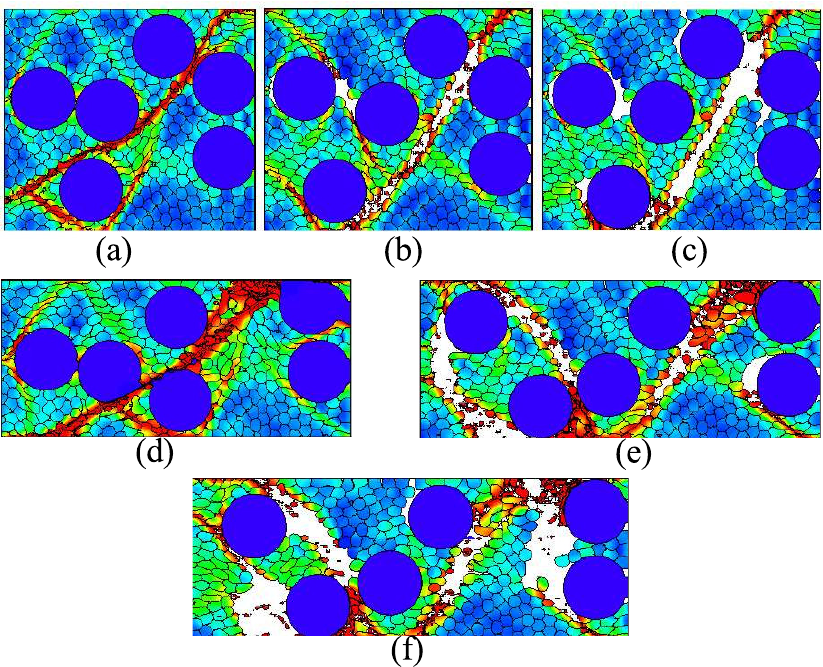}
\caption{\label{fig:13}  (Color online) Role of different conditions of confinement on shear localization and subsequent fracture.  Three deformed mesostructures at 0.3 and 0.5 global strains with various confinement conditions are presented. All three samples have boundary conditions described in Fig. \ref{fig:7} (b) with a constant 10 m/s downward velocity imposed on the top boundary. Each sample differs only in the imposed horizontal velocity on the side boundaries (linear ramp as vertical strain goes from 0 to 0.5) in the following way:\\
(a),(d) A linear ramp from 0 to 7.20 m/s. Corresponding to a final horizontal strain equal to 0.58. This produces the representative element to keep the same geometric proportions as in the experiments. Sample shown at 0.3 (a) and 0.5 (d) vertical strain.\\
(b),(e) A linear ramp from 0 to 12.24 m/s. This corresponds to a horizontal strain 1.0. Sample shown at 0.3 (b) and 0.5 (e) vertical strain.\\
(c),(f) A linear ramp from 0 to 14.40 m/s. This corresponds to a horizontal strain of 1.16. Sample shown at 0.3 (c) and 0.5 (f) vertical strain.  }
\end{figure}

Previous simulations examined small representative elements corresponding to a material on, or very near to, the outer edge of the experimental sample. A small representative element near the central region of the sample interacts with surrounding material which resists movement of material leaving the representative element. To account for this “confinement” from the surrounding material, kinematic boundary conditions using a ramped normal velocity were imposed on the side. Multiple velocity ramps were explored in an attempt to model the large variance in the local inhomogeneity within the sample, resulting in various local “confinements”. The first ramp velocity tested kept the global Poisson ratio of the small representative sample geometrically consistent with that of the experimental sample. Two additional constrained boundary conditions were tested corresponding to 70\% and 100\% increased horizontal expansion in comparison with the geometrically constrained case. The vertical strain rate was kept identical for all samples. Fig. \ref{fig:13} shows the simulation results at 0.30 and 0.50 global strains respectively for the samples with the constrained side boundary conditions. Damage is plotted to highlight the shear instabilities. 

Shear bands developed in all samples at approximately 0.25 global strain. However, the location and number of shear bands differed in each case. Despite these differences, all the samples share a similar characteristic: areas of heavily deformed Al particles near the W particles, resulting in local damage accumulation and subsequent shear band formation, while material further away from the W particles is relatively undeformed. This trend seems to be independent of the imposed confinement conditions. Also, samples in Fig. \ref{fig:13} (b) and (c) exhibit cracks that form between the W particles. The crack locations are very similar to the location of cracks in the large experimental sample (see Fig. \ref{fig:3} (b)). The cracks span between the W particles in the areas of the heavily deformed Al. In the simulations, the W particles move, squeezing the Al particles between the nearby W particles, acting as “anvils” that facilitate the heterogeneous deformation of the Al particles. A similar effect may occur in the large experimental sample where the Al flows around the W fibers due to the three-dimensional cage like network that the packed network of W fibers create. A similar effect of rigid and heavy W particles is instrumental also in the shock loading of corresponding mixtures allowing tailored redistribution of internal energy between components.\cite{Cai2,Herbold}

\subsection{The role of the constitutive behavior of Al}
\begin{figure}
\includegraphics{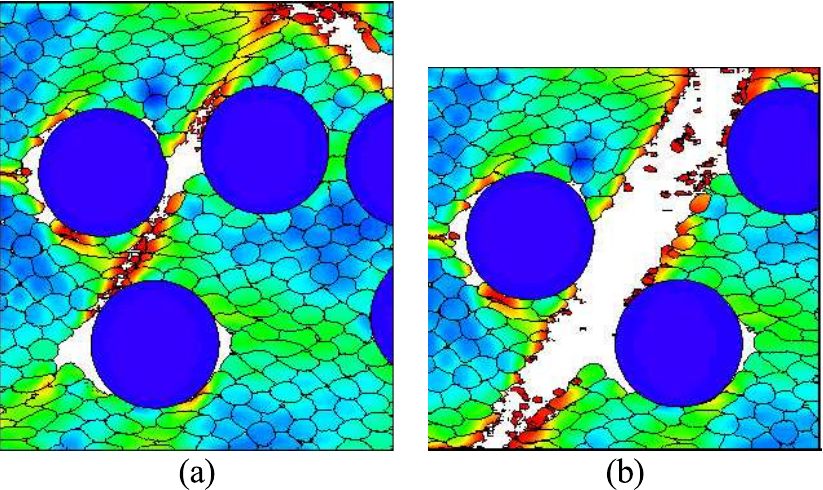}
\caption{\label{fig:14}  (Color online) Effect of the initial yield strength on the shear instability in porous granular composites. The initial yield strength of the Al is reduced to that of a softer Al (20 MPa). The sample is presented with  global strains of 0.30 (a) and 0.40 (b). The damage is plotted to highlight the shear bands.   }
\end{figure}

The constitutive behavior of the Al particles were modified to explore how they influenced the shear instability and the shear band formation in the bonded samples (CIPed plus HIPed samples in experiments). Simulations were carried out with the initial mesostructure shown in Fig. \ref{fig:5} and the boundary conditions in Fig. 7 (a) with a constant 10 m/s downward velocity on the top boundary. The initial yield stress of Al was reduced to a very low level 20 MPa and the results are shown in Fig. \ref{fig:14}. The sensitivity of shear band development to other parameters in the Johnson-Cook model was also explored, however, these simulations showed that there were only slight changes in the material response due to these alterations and results are not presented in this paper.  

The reduction of the initial yield stress in the material from that of Al 6061-T6 (324 MPa) to that of a much softer Al (20 MPa) caused a large change in sample response (compare to the results in Fig. \ref{fig:10} (a) and (b)). The Al particles in Fig. \ref{fig:14} experiences greater plastic flow and deformation in all areas of the sample, especially in the areas around the W particles. Due to the softening of the Al matrix, the development of the global shear band was delayed, allowing for more bulk-distributed plastic flow of the Al. The reduction of initial yield stress in the Al resulted in a shear band developing in a different location. This is likely caused by the greater movement of W particles in the softer Al altering the arrangement of the W altering the mesostructure at the earlier stages of sample deformation. 
\section{Conclusions}
Dynamic experiments with Al-W granular/porous composites and numerical simulations revealed the characteristics that had the greatest effects on shear instability and shear band formation. It was shown that in simulated CIPed only (unbonded) samples the Al and W particles rearranged themselves during the dynamic deformation to effectively block shear localization. This resulted in the subsequent bulk disintegration of the sample in agreement with areas near the outer surface of the samples used in the experiments.

All CIPed+HIPed (bonded) samples exhibited shear localization and shear band formation. The shear bands nucleated during the initial stages of the deformation in Al surrounding the W particles and spread to the nearby W particles at angles close to 45 degrees. The shear band is kinked by W particles causing the shear band path to deviate from the ideal 45 degree angle path dictated by global geometry. In simulations with relatively larger W particles, the path of the shear band was influenced to a greater degree than simulations with the similar sized W and Al particles due to increased heterogeneity of the sample. It was also shown that variations in the initial arrangements of the W particles were the main drivers determining where the global shear bands formed in the sample. Numerical calculations and experiments revealed that the mechanism of shear localization in granular composites is due to a localized high strain flow of Al around rigid W particles, causing local damage accumulation and a subsequent growth of the meso/macro shear bands/cracks.  

A variety of constraint boundary conditions were examined to represent the heterogeneous nature of the internal structure in the global sample. Each simulation showed shear localization occurred between the nearby W particles while leaving areas away from the W particles relatively undeformed. This result is supported by the microstructural features observed in the experimental sample.
	
Finally, the role of the constitutive behavior in the Al was examined in numerical calculations. The results showed that a significant reduction of the initial yield stress from Al 6061-T6 (324 MPa) to a softer Al (20 MPa) increased the amount of bulk distributed damage and plastic strain in the sample in addition to altering the shear band location.   
\section{Acknowledgements}
The support for this project provided by the Office of Naval Research Multidisciplinary University Research Initiative Award N00014-07-1-0740.

\providecommand{\noopsort}[1]{}\providecommand{\singleletter}[1]{#1}%

\end{document}